\documentclass[fleqn,10pt]{wlscirep}
\usepackage[utf8]{inputenc}
\usepackage[T1]{fontenc}
\usepackage{bbm}
\usepackage{algorithm}
\usepackage[noend]{algpseudocode}

\newtheorem{definition}{Definition}
\title{WHAT COMES NEXT? RESPONSE TIMES ARE AFFECTED BY MISPREDICTIONS IN A STOCHASTIC GAME \\ \vspace{0.25cm} \Large{ \textit{Dedicated to the memory of Antonio Galves} } }

\author[1]{Paulo Roberto Cabral-Passos}
\author[2]{Antonio Galves}
\author[3]{Jesus Enrique Garcia}
\author[4,*]{Claudia D. Vargas}
\affil[1]{Departamento de Física da Faculdade de Filosofia, Ciências e Letras de Ribeirão Preto, Universidade de São Paulo, Ribeirão Preto, Brazil
}
\affil[2]{Instituto de Matemática e Estatística, Universidade de São Paulo, São Paulo, Brazil}
\affil[3]{Instituto de Matemática, Estatística e Computação Científica, Universidade Estadual de Campinas, Campinas, Brazil}
\affil[4]{Instituto de Biofísica Carlos Chagas Filho, Universidade Federal do Rio de Janeiro, Rio de Janeiro, Brazil}

\affil[*]{cdvargas@biof.ufrj.br}

\begin{abstract}

Acting as a goalkeeper in a video-game, a participant is asked to predict the successive choices of the penalty taker. The sequence of choices of the penalty taker is generated by a stochastic chain with memory of variable length. It has been conjectured that the probability distribution of the response times is a function of the specific sequence of past choices governing the algorithm used by the penalty taker to make his choice at each step. We found empirical evidence that besides this dependence, the distribution of the response times depends also on the success or failure of the previous prediction made by the participant. Moreover, we found statistical evidence that this dependence propagates up to two steps forward after the prediction failure.

\end{abstract}
\begin{document}

\flushbottom
\maketitle

\thispagestyle{empty}

\section*{Introduction}

More than a century ago, Helmholtz\cite{Helmholtz:1867} conjectured that the human brain is able to detect statistical regularities in a sequence of events. Since then, psychophysiological measurements have been employed to study this conjecture \cite{Nissen:1987, Hunt:2001, Visser:2007, Baldwin:2008, Dehaene:2015, Frost:2015, Kahn:2018,Lange:2018,Wang:2017,Wang:2017b}. Recently, the classical conjecture proposed by Helmholtz was revisited using a new probabilistic framework \cite{Duarte:2019,Hernandez:2021}. In Duarte et al.\cite{Duarte:2019}, the relationship between a sequence of auditory stimuli and the sequence of EEG segments recorded during the exposure to these stimuli was modelled using sequences of random objects driven by a stochastic chain with memory of variable length. Using the framework introduced by Duarte et al.\cite{Duarte:2019}, Hernandez et al. \cite{Hernandez:2021} provided statistical evidence that the probability distribution of the EEG segments depended on the smallest sequence of past auditory stimuli governing the choice of the next auditory stimulus. Following Rissanen\cite{Rissanen:1983}, the smallest sequence of past stimuli governing the probabilistic choice of the next stimulus is called a \textit{context}. Moreover, Rissanen observed that the set of all contexts describing a stochastic chain can be described as the set of leaves of a rooted and labeled tree. For this reason, from now on we will refer to the set of contexts as a \textit{context tree}. It is natural to conjecture that this dependence on the context tree proposed in Duarte et al \cite{Duarte:2019} and employed in Hernandez et al. \cite{Hernandez:2021} would also occur at a behavioral level. This is the starting point of the present work. 

To address this issue, we developed a video-game called the Goalkeeper Game \cite{gkgame, Stern:2020, Hernandez:2023}. In the Goalkeeper game, the penalty taker has three available action choices: kick to the left, to the center, or to the right side of the goal. The sequence of choices of the penalty taker is generated by a stochastic chain with memory of variable length whose dependence on the past is described by a context tree. Acting as a goalkeeper, the participant must predict at each step which will be the next choice of the penalty taker. The participant is instructed to save the maximum number of balls. Response times of the participant are recorded at each trial.  After the trial, a feedback video indicates the goalkeeper's success or failure. The Goalkeeper game offers an opportunity to simulate an environment in which prediction of an upcoming sensorimotor event is necessary and its product is expressed as a prediction success or failure. 

In the present framework, we look at the relationship between the probability distribution of response times and the sequence of contexts displayed by the successive choices of the penalty taker. We provide statistical evidence that, besides the dependence on the contexts, the probability distribution of the response times depends also on the success or failure of the previous
predictions made by the goalkeeper.

\section*{Methods}

\subsection*{Experimental protocol}

 The following experimental protocol was approved by the Ethics Committee of the Institute of Neurology Deolindo Couto at the Federal University of Rio de Janeiro (CAEE: 58047016.6.1001.5261). Twenty-two right-handed participants (14 females) were invited to play remotely the online version of the Goalkeeper Game \cite{gkgame}. In this game version, the participant assumes the role of a goalkeeper in a sequence of 1000 penalty trials. The directions of choice were towards left, center and right. For simplicity, we indicate these directions by the numerical symbols 0, 1 and 2, respectively (Fig. \ref{fig:painel1}A). At each trial, acting as a goalkeeper, the participant chooses where to jump to save the kick by pressing the left arrow key with the right index finger (0), the down arrow key with the right middle finger (1), or the right arrow key with the right ring finger (2). Two rest intervals were placed along the 1000 trials, the first after the trial 334 and the second after the trial 668. The mean and standard deviation of the first and second rest intervals were $54 \pm 55 sec.$ and $50 \pm 40 sec.$, respectively. The penalty taker choices were not influenced by the previous choices of the goalkeeper. Besides, the goalkeeper was told to take his/her time to make his/her decision and to resume the game after rest intervals. In each trial, the penalty kick took place only after the participant has conveyed his decision by pressing a button.

 The sequence of kicks was generated by a stochastic chain with memory of variable length whose dependence on the past is described by a context tree $\tau$.  Let $p$ be the family of transition  probabilities indexed by the contexts in $\tau$,  governing the successive choices made by the penalty taker given the corresponding context. The pair $(\tau, p)$ will be called a \textit{ probabilistic context tree}  \cite{smselection}. 

 The probabilistic context tree  $(\tau,p)$ used in our experimental protocol is described in Figure \ref{fig:painel1}B, which also  shows an example of a sequence generated by $(\tau, p)$. This stochastic sequence can also be described as a concatenation of successive choices of the sequence  $0 * 1$, where at each repetition the symbol $*$ is replaced either by $2$, with probability $p = 0.7$, or by $1$ with probability $1-p$, independently of the previous choices.

\subsection*{Estimating a context tree from the sequence of response times}

Let $(X_{n}: n=1,\ldots, 1000)$ and $(Y_{n}: n=1,\ldots, 1000)$ be, respectively, the sequences of directions chosen by the penalty taker and by the goalkeeper during the game. Both $X_n$ and $Y_n$ belong to the set of  possible directions $A=\{0,1,2\}$. We say that the $n$-th prediction is correct when $X_{n} = Y_{n} $.  Let also $(T_{n}: n=1,\ldots, 1000)$ be the corresponding sequence of response times of the goalkeeper, see Figure \ref{fig:painel1}. Given a sequence $w$, $l(w)$ is the length of $w$.

The following algorithm extends Rissanen's \textit{Context} algorithm to sequences of real numbers driven by a probabilistic context tree. In the presentation of the algorithm, the word \textit{list} is used in the sense it has in the Python language.  

The algorithm uses the {\it reverse lexicographical order} to arrange the sequences. 

\begin{definition}
The reverse lexicographical order between sequences of length $K$ is defined as follows: 
$(u_{-K},\cdots ,u_{-1}) < (v_{-K},\cdots,v_{-1})$ if either $u_{-1} < v_{-1}$, or there exists $2 \leq j \leq K$ such that $(u_{-j+1},\cdots ,u_{-1})=v_{-j+1},\cdots ,v_{-1}$ and $u_{-j} < v_{-j}$.
\end{definition}

\subsection*{Algorithm Steps}

\begin{enumerate}
    \item[] \textbf{Initialization}: The algorithm begins by initializing an empty context tree $\hat{\tau}$ and a list $C$, containing all the sequences of length $K$ appearing in the sample.
    \item[] \textbf{Iterative Process}: The algorithm proceeds in an iterative manner until the set $C$ is empty. Within each iteration:
    
    \begin{enumerate}
        \item The first sequence $w$ in the list $C$ is selected. 
        
        \item A new list $F(w)$ is formed. This list contains all the sequences appearing in the sample, that can be obtained by appending, as first element,  a symbol from the alphabet $A$ to the sequence $(w_{-l(w)+1},\cdots ,w_{-1})$.
        
 \item If $F(w)\subseteq C$, the Kolmogorov-Smirnov test is used to decide if the distribution of the response times corresponding to the members of the list  $F(w)$
 are the same. 
 
\begin{enumerate}
\item If the Kolmogorov-Smirnov test rejects the equality of distributions, then the sequences in $F(w)$ are added to $\hat{\tau}$ and deleted from the List $C.$ 
\item Otherwise,  the sequences in $F(w)$ are deleted from the list $C$ and the sequence $(w_{-l(w)+1},\cdots ,w_{-1})$ is added to the end of $C.$ 
\item In the case of $F(w)=\left\{w\right\},$  $w$ is deleted from $C$ and $(w_{-l(w)+1},\cdots ,w_{-1})$ is added to the end of $C.$ 
\end{enumerate}

\item If $F(w)\not\subseteq C$, the sequences in $F(w)\cap C$ are deleted from the list $C$ and  added to $\hat{\tau}.$ 
   
    \end{enumerate}

    \item \textbf{Output}: Once all iterations are complete and the  list $C$ is empty, the algorithm outputs the constructed context tree $\hat{\tau}$.
\end{enumerate}

\subsection*{Epochs and mode context tree}
To  access the evolution of the context trees across time, the sequence of response times per participant was divided into three epochs, separated in accordance with the position of rest intervals in the sequence of trials.  The first epoch goes from $1$ to $334$; the second epoch goes from $335$ to $668$, and the third epoch goes from $669$ to $1000$. Context trees by epoch and participant were estimated using the algorithm described above. 
For each  epoch, the set of context trees retrieved from the data collected for all the participants was then summarized  through a \textit{mode context tree}. The mode context tree contains all and only the contexts which appear more frequently across participants, see Figure 3 in Hernández et al.\cite{Hernandez:2021} .

\subsection*{Response time comparison according to the result of previous predictions}

Given a sequence $w=(w_{-k}, \ldots, w_{-1})$, let $n_{(1,w)},n_{(2,w)}\ldots $ be the  successive steps ending in an occurrence of $w$. Namely, 

$$n_{(1,w)}=\min\{n\geq k: ~X_{n-k+1}=w_{-k},\cdots , X_{n}=w_{-1}\},$$
and for $j>1$
$$n_{(j,w)}=\min\{n > n_{(j-1,w)}: ~   X_{n-k+1}=w_{-k},\cdots , X_{n}=w_{-1}\}.$$

Given a sequence $w=(w_{-k}, \ldots ,w_{-1})$, let $n_{(1,w,s)},n_{(2,w,s)}\ldots $ be the  successive steps ending in an occurrence of $w$ after a correct prediction following a $0$ in the sequence. In the same way, let $n_{(1,w,f)},n_{(2,w,f)}\ldots $ be the  successive steps ending in an occurrence of $w$ after an incorrect prediction following a $0$ in the sequence.

Let $N(w,f)$ be the total number of  occurrences of $w$ after a correct prediction following a $0$ and $N(w,s)$ the  total number of  occurrences of $w$ after a failure in prediction following a $0$.

Let  $T_i^{w,s}=T_{n_{(i,w,s)}+1},$ for $i=1,\cdots, N(w,s)$ and $T_i^{w,f}=T_{n_{(i,w,f)}+1},$ for $i=1,\cdots, N(w,f)$ be the set of response times after a correct  and an incorrect prediction, respectively, following a $0$.

Let $\bar{T}^{(w,s)}$ and  $\bar{T}^{(w,f)}$ be the sample mean response time  after a correct prediction and an incorrect prediction following a $0$, respectively.

\begin{figure}[ht]
    \centering
	\includegraphics[scale=0.055]{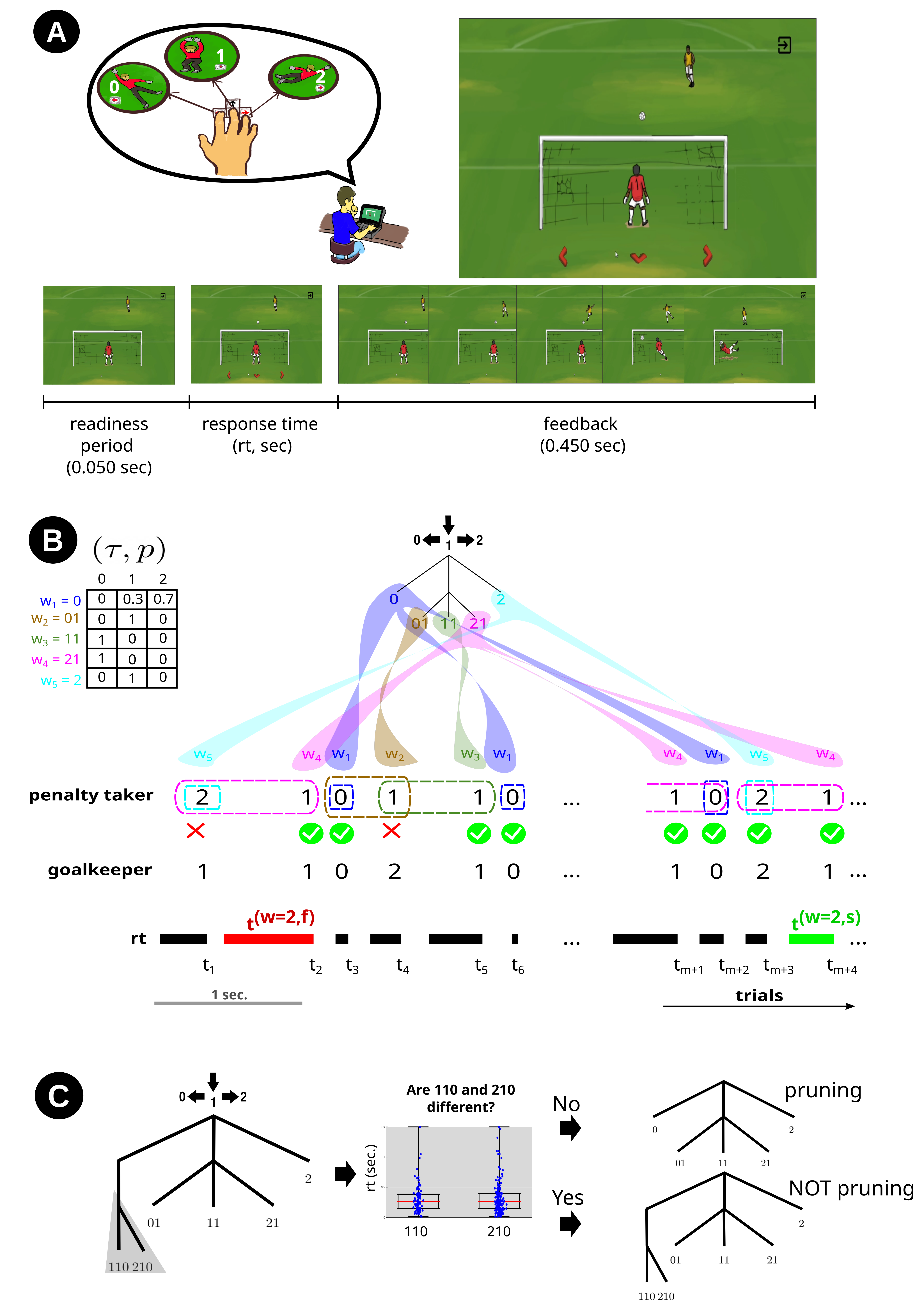}
\end{figure}

\vfill
\clearpage

\begin{figure}
    \centering
    \caption{The Goalkeeper Game experiment. A) At each trial, acting as a goalkeeper the participant chooses where to jump to save the kick by pressing the left arrow key with the right index finger (0), the down arrow key with the right middle finger (1) or the right arrow key with the right ring finger (2). The top right picture shows the screen at the moment of the decision. The bottom diagram shows the timeline of a trial with the duration of each segment. Each trial starts with a readiness period. Then, arrows appear at the bottom of the screen, indicating that the participant is allowed to inform his decision by pressing one of the buttons indicated in the balloon. The time from the appearance of the arrows till the button press is defined as the response time (rt). Immediately after the button press, the feedback is presented as an animation depicting the kicker's choice. B) Top left, the context tree probability table used for generating the sequence of kicks of the penalty taker. On the top right, the set of contexts $\tau$ is represented as a labeled and rooted tree. A different color is attributed to each context. An example of a sequence of penalty taker choices and the corresponding sequence of predictions of the goal\-keeper are shown at the bottom of the picture. The green check mark indicates a successful prediction, while the red cross ($\times$) indicates a prediction failure. On the bottom, successive response times ($t$) are shown as horizontal bars in which the width represents the response time duration in seconds from a real participant. Here, $m$ corresponds to a positive integer used to illustrate the shift to trials far from those before the three dots. Response times in the context $2$, $t^{(w=2)}$, are highlighted in red after an incorrect prediction, $t^{(w = 2,f)}$, and in green after a correct prediction, $t^{(w = 2,s)}$. C) An example of the pruning procedure used to estimate context trees from response times of a given participant. If the law of response times of at least two leaves of a branch is statistically different, the branch is presented in the estimated tree. The picture illustrates this procedure for the pair of leaves $110$ and $210$.}
    \label{fig:painel1}
\end{figure}

\section*{Results}

\begin{figure}[h]
    \centering
    \includegraphics[scale=0.35]{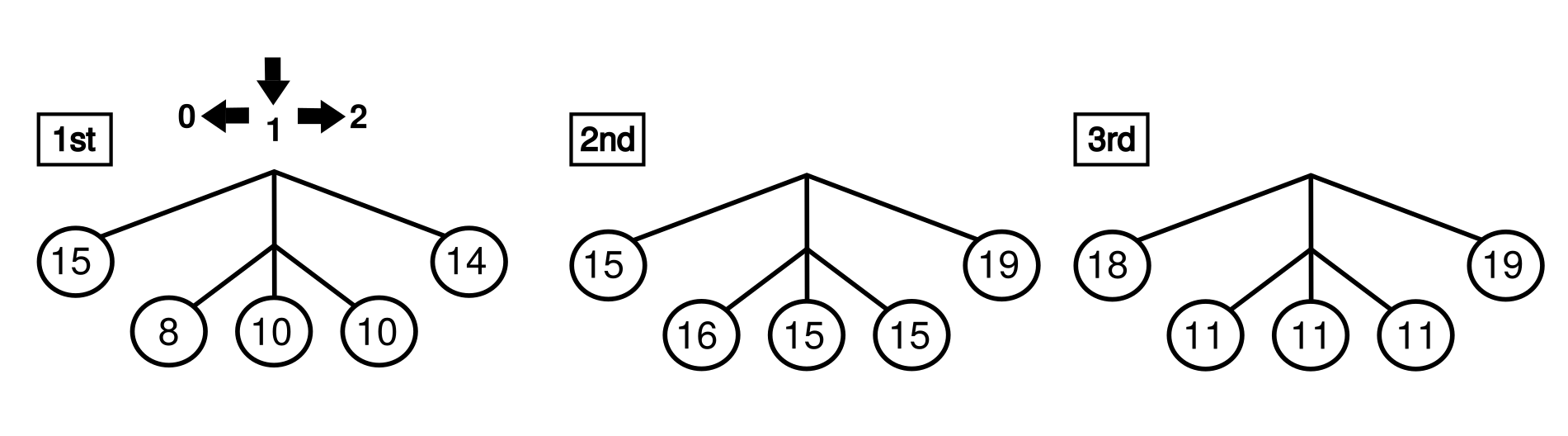}
    \caption{Mode context trees for each epoch. The numbers inside the circles indicate the number of participants (n=22) for which the corresponding context was present in the estimated trees of that epoch.}
    \label{fig:painel2}
\end{figure}

\indent Response times were employed to estimate context trees per participant and per epoch (Figure \ref{fig:painel2}). For all epochs, the mode context tree was the same as the context tree used by the penalty taker to generate the sequence of kicks. Moreover, the number of participants who correctly identified contexts 0 and 2 increased from the first to the third epoch. Curiously, the correct identification of contexts ending in 1 increased from the first to the second epoch, but diminished from the second to the third epoch. 
Since the sequence of kicks consists in a repetition of $0 * 1$ with $*$ taking the value of $2$ with probability $p = 0.7$ and $1$ with probability $1-p$, we reasoned that the contexts $01$, $11$ and $21$ might be affected by the congruence between the participants choices and those of the penalty taker. 

To evaluate the influence of past predictions over response times in a given context, response times were divided into two sub-samples (see Figure \ref{fig:painel1}B). $T_1^{(w,s)}, T_2^{(w,s)}, \ldots$ indicate the response times in $w$ given that the participant successfully predicted the choice of the penalty taker the last time the context $0$ took place. Similarly, $T_1^{(w,f)}, T_2^{(w,f)}, \ldots$ indicate the response times in $w$ given that the participant failed to predict the choice of the penalty taker the last time the context $0$ took place. This was done because the participant who has learned the regularities of the sequence would only fail to predict the penalty taker's choice in that context. The mean values of the response times for each participant, context and sub-sample can be found in supplementary table 2.

Figure \ref{fig:painel3} shows the distributions of response times after correct and incorrect predictions at the last time the context $0$ took place, that is, $T_1^{(w,s)}, T_2^{(w,s)}, \ldots$ and $T_1^{(w,f)}, T_2^{(w,f)}, \ldots$, for one participant. To test if the mean values $\bar{T}^{(w,f)}$ were higher than the mean values $\bar{T}^{(w,s)}$, the difference $\bar{T}^{(w,f)} - \bar{T}^{(w,s)}$ was calculated for each context and participant using the trimmed mean\cite{Hill:1982}. A one-tailed Wilcoxon signed-rank test showed that these differences were significantly different from zero for the contexts $01$, $2$ and $21$.  The test indicated that the mean response times for context $w = 2$ were smaller after successful predictions compared to after prediction failures ($Z = 4.09$, $p = 2.1 \times 10^{-5}$). This was also true for context $21$, which occurs one step further in the sequence, however, with a less pronounced effect ($Z = 2.434$, $p = 0.007$). On the other hand, for context $w = 01$, after successful predictions the mean response times were larger than after prediction failures ($Z = -2.467$, $p = 0.006$). For the context $11$ the effect was only close to statistical significance ($Z = -1.363$, $p = 0.086$), but it is important to highlight that $11$ is the less frequent context of the sequence. Finally, the context $0$ presented no significant difference from zero ($Z = 0.47$, $p = 0.637$). Taken together, these results indicate that the distribution of response times changes as a function of the result of previous predictions.

\begin{figure}
    \centering
    \includegraphics[scale=0.25]{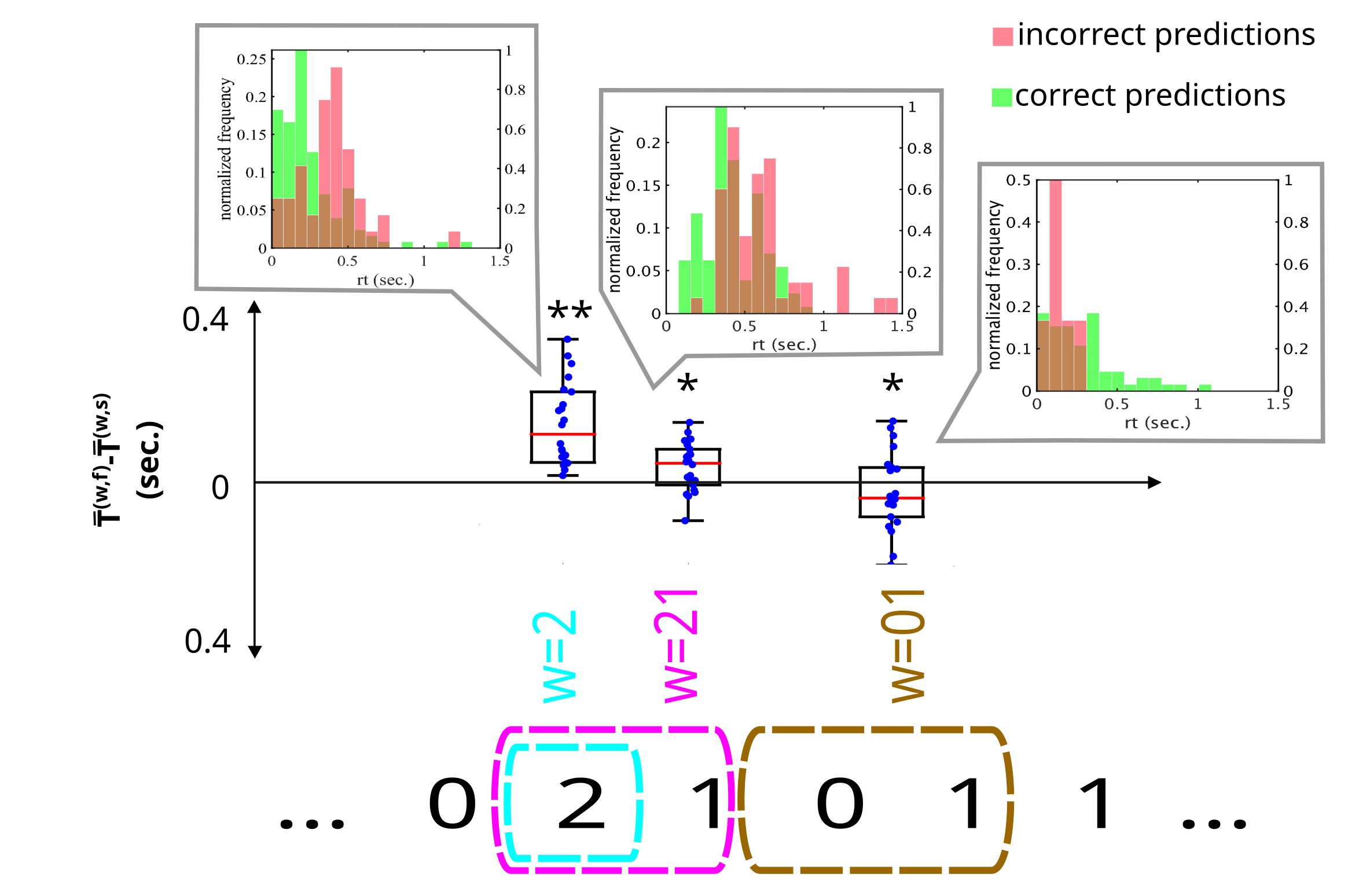}
    \caption{Difference between the mean values $\bar{T}^{(w,f)}$ and $\bar{T}^{(w,s)}$ for the contexts $2$, $21$ and $01$ for which the trimmed mean was statistically significant according to the Wilcoxon signed-rank test. $*$ indicates a $p < 0.01$ and $**$ indicates a $p = 2.1 \times 10^{-5}$. Each blue dot shows the paired difference per participant. On top of each distribution, superimposed histograms of the response times of one participant illustrate the differences between $\bar{T}^{(w,s)}$ in green and $\bar{T}^{(w,f)}$ in red. The sequence of contexts governing the choices of the penalty taker is presented at the  bottom.}
    \label{fig:painel3}
\end{figure}

The Kruskal-Wallis test ($\chi^{2} = 2.55$, $df = 2$, $p = 0.28$) indicated no difference between the response times for different fingers.

\section*{Discussion}

Response times associated to a stochastic sequence of events were investigated using the Goalkeeper Game. The sequence of choices of the penalty taker was generated by a stochastic chain with memory of variable length and can be expressed as a repetition of $0~*~1$, in which the middle position $*$ is replaced by a $2$ with probability $p=0.7$ and by $1$ with probability $1-p$, independently of the goalkeeper's choices. The statistical analysis of the data provided the following results.

First of all, we successfully retrieved the context tree used by the penalty taker from the goalkeeper's response time. This supports the conjecture that the probability distributions of the goalkeeper's response times depend on the contexts governing the choices of the penalty taker at each step.  
Previous studies reported that response times are affected by the stochasticity of the sequence  of stimuli \cite{Visser:2007, Wang:2017, Wang:2017b, Kahn:2018}.    To the best of our knowledge, this is the first study using response times in which the structure of the sequence of random stimuli is retrieved from the participant's response times. 

Moreover, we found that the number of participants from whose response times allowed to correctly retrieve the penalty taker tree increased from the first to the second epoch of the game. More precisely, in the second epoch, the mode context tree deduced from the response times of a large majority of participants (15 out of 22) coincided with the context tree used to generate the sequence of choices of the penalty taker. Surprisingly, the number of participants whose response times allowed to correctly identify the contexts 01, 11 and 21 decreased from the second to the third epoch (only 11 out of 22 participants). 
This suggested that an extra factor could be at play. In fact, besides being governed by the context, our statistical analysis provided evidence that response times were also affected by the result of previous predictions. 

Response times in a same context depended on the result of previous predictions and this dependence propagated up to two steps forward. This was shown for the contexts $2$, $21$ and $01$, for which different mean response times were identified according to the success or failure of the prediction made by the goalkeeper in the previous occurrence of context $0$. Smaller response times were found after successful as compared to unsuccessful predictions for the contexts $2$ and $21$. On the other hand, larger response times in the context $01$ were found after successful as compared to unsuccessful predictions. 

The slowing of response times after unsuccessful predictions for the contexts $2$ and $21$ finds a parallel in studies using serial reaction time paradigms \cite{Rabbitt:1966, Rabbitt:1977,Gehring:2001,Jentzsch:2009,Eichele:2010,Danielmeier:2011,Danielmeier:2011b, Braem:2015}. As a matter of fact, the slowing of response times after an immediate mistake in serial reaction time paradigms has been associated with a temporary inhibition of the primary motor cortex, as measured by EEG and fMRI \cite{Danielmeier:2011,Danielmeier:2011b}. In the present study, we show that for the context $01$ the subsequent response times also increase after successful predictions, suggesting that other factors may play a role in the response times slowing, see for instance \cite{Wessel:2018}.

In conclusion, we were able to retrieve the statistical regularities from a sequence of response times by applying the Context Tree algorithm\cite{Hernandez:2021, Hernandez:2023}. This was done by modelling the relationship between the sequence of response times of a given participant and the stochastic sequence of choices of a penalty taker. With this approach, we found that response times are influenced both by contexts and by the results of previous predictions. The Goalkeeper game gives the opportunity to simulate an environment in which prediction is necessary and its product is verifiable. With this information, it is possible to understand new aspects of motor behaviour. 

\bibliography{ref}

\begin{thebibliography}{10}
\urlstyle{rm}
\expandafter\ifx\csname url\endcsname\relax
  \def\url#1{\texttt{#1}}\fi
\expandafter\ifx\csname urlprefix\endcsname\relax\def\urlprefix{URL }\fi
\expandafter\ifx\csname doiprefix\endcsname\relax\def\doiprefix{DOI: }\fi
\providecommand{\bibinfo}[2]{#2}
\providecommand{\eprint}[2][]{\url{#2}}

\bibitem{Helmholtz:1867}
\bibinfo{author}{Helmholtz, H.~V.}
\newblock \bibinfo{title}{Handbuch der physiologischen optik}
  (\bibinfo{year}{1867}).

\bibitem{Nissen:1987}
\bibinfo{author}{Nissen, M.~J.} \& \bibinfo{author}{Bullemer, P.}
\newblock \bibinfo{journal}{\bibinfo{title}{Attentional requirements of
  learning: Evidence from performance measures}}.
\newblock {\emph{\JournalTitle{Cognitive Psychology}}}
  \textbf{\bibinfo{volume}{19}}, \bibinfo{pages}{1--32},
  \doiprefix\url{https://doi.org/10.1016/0010-0285(87)90002-8}
  (\bibinfo{year}{1987}).

\bibitem{Hunt:2001}
\bibinfo{author}{Hunt, R.~H.} \& \bibinfo{author}{Aslin, R.~N.}
\newblock \bibinfo{journal}{\bibinfo{title}{Statistical learning in a serial
  reaction time task: Access to separable statistical cues by individual
  learners}}.
\newblock {\emph{\JournalTitle{Journal of Experimental Psychology. General}}}
  \textbf{\bibinfo{volume}{130}}, \bibinfo{pages}{658--680},
  \doiprefix\url{https://doi.org/10.1037//0096-3445.130.4.658}
  (\bibinfo{year}{2001}).

\bibitem{Visser:2007}
\bibinfo{author}{Visser, I.}, \bibinfo{author}{Raijmakers, M. E.~J.} \&
  \bibinfo{author}{Molenaar, P. C.~M.}
\newblock \bibinfo{journal}{\bibinfo{title}{Characterizing sequence knowledge
  using online measures and hidden markov models}}.
\newblock {\emph{\JournalTitle{Memory and Cognition}}}
  \textbf{\bibinfo{volume}{35}}, \bibinfo{pages}{1502--1517},
  \doiprefix\url{https://doi.org/10.3758/BF03193619} (\bibinfo{year}{2007}).

\bibitem{Baldwin:2008}
\bibinfo{author}{Baldwin, D.}, \bibinfo{author}{Andersson, A.},
  \bibinfo{author}{Saffran, J.} \& \bibinfo{author}{Meyer, M.}
\newblock \bibinfo{journal}{\bibinfo{title}{Segmenting dynamic human action via
  statistical structure}}.
\newblock {\emph{\JournalTitle{Cognition}}} \textbf{\bibinfo{volume}{106}},
  \bibinfo{pages}{1382--1407},
  \doiprefix\url{https://doi.org/10.1016/j.cognition.2007.07.005}
  (\bibinfo{year}{2008}).

\bibitem{Dehaene:2015}
\bibinfo{author}{Dehaene, S.}, \bibinfo{author}{Meyniel, F.},
  \bibinfo{author}{Wacongne, C.}, \bibinfo{author}{Wang, L.} \&
  \bibinfo{author}{Pallier, C.}
\newblock \bibinfo{journal}{\bibinfo{title}{The neural representation of
  sequences: From transition probabilities to algebraic patterns and linguistic
  trees}}.
\newblock {\emph{\JournalTitle{Neuron}}} \textbf{\bibinfo{volume}{88}},
  \bibinfo{pages}{2--19},
  \doiprefix\url{https://doi.org/10.1016/j.neuron.2015.09.019}
  (\bibinfo{year}{2015}).

\bibitem{Frost:2015}
\bibinfo{author}{Frost, R.}, \bibinfo{author}{Armstrong, B.~C.},
  \bibinfo{author}{Siegelman, N.} \& \bibinfo{author}{Christiansen, M.~H.}
\newblock \bibinfo{journal}{\bibinfo{title}{Domain generality versus modality
  specificity: the paradox of statistical learning}}.
\newblock {\emph{\JournalTitle{Trends in Cognitive Sciences}}}
  \textbf{\bibinfo{volume}{19}}, \bibinfo{pages}{117--125},
  \doiprefix\url{https://doi.org/10.1016/j.tics.2014.12.010}
  (\bibinfo{year}{2015}).

\bibitem{Kahn:2018}
\bibinfo{author}{Kahn, A.~E.}, \bibinfo{author}{Karuza, E.~A.},
  \bibinfo{author}{Vettel, J.~M.} \& \bibinfo{author}{Bassett, D.~S.}
\newblock \bibinfo{journal}{\bibinfo{title}{Network constraints on learnability
  of probabilistic motor sequences}}.
\newblock {\emph{\JournalTitle{Nature Human Behaviour}}}
  \textbf{\bibinfo{volume}{2}}, \bibinfo{pages}{936--947},
  \doiprefix\url{https://doi.org/10.1038/s41562-018-0463-8}
  (\bibinfo{year}{2018}).

\bibitem{Lange:2018}
\bibinfo{author}{Lange, F. P.~D.}, \bibinfo{author}{Heilbron, M.} \&
  \bibinfo{author}{Kok, P.}
\newblock \bibinfo{journal}{\bibinfo{title}{How do expectations shape
  perception?}}
\newblock {\emph{\JournalTitle{Trends in Cognitive Sciences}}}
  \textbf{\bibinfo{volume}{22}}, \bibinfo{pages}{764--779},
  \doiprefix\url{https://doi.org/10.1016/j.tics.2018.06.002}
  (\bibinfo{year}{2018}).

\bibitem{Wang:2017}
\bibinfo{author}{Wang, R.}, \bibinfo{author}{Shen, Y.}, \bibinfo{author}{Tino,
  P.}, \bibinfo{author}{Welchman, A.~E.} \& \bibinfo{author}{Kourtzi, Z.}
\newblock \bibinfo{journal}{\bibinfo{title}{Learning predictive statistics:
  Strategies and brain mechanisms}}.
\newblock {\emph{\JournalTitle{The Journal of Neuroscience}}}
  \textbf{\bibinfo{volume}{37}}, \bibinfo{pages}{8412--8427},
  \doiprefix\url{https://doi.org/10.1523/JNEUROSCI.0144-17.2017}
  (\bibinfo{year}{2017}).

\bibitem{Wang:2017b}
\bibinfo{author}{Wang, R.}, \bibinfo{author}{Shen, Y.}, \bibinfo{author}{Tino,
  P.}, \bibinfo{author}{Welchman, A.~E.} \& \bibinfo{author}{Kourtzi, Z.}
\newblock \bibinfo{journal}{\bibinfo{title}{Learning predictive statistics from
  temporal sequences: Dynamics and strategies}}.
\newblock {\emph{\JournalTitle{Journal of Vision}}}
  \textbf{\bibinfo{volume}{17}}, \bibinfo{pages}{1},
  \doiprefix\url{https://doi.org/10.1167/17.12.1} (\bibinfo{year}{2017}).

\bibitem{Duarte:2019}
\bibinfo{author}{Duarte, A.}, \bibinfo{author}{Fraiman, R.},
  \bibinfo{author}{Galves, A.}, \bibinfo{author}{Ost, G.} \&
  \bibinfo{author}{Vargas, C.~D.}
\newblock \bibinfo{journal}{\bibinfo{title}{Retrieving a context tree from eeg
  data}}.
\newblock {\emph{\JournalTitle{Mathematics}}} \textbf{\bibinfo{volume}{7}},
  \bibinfo{pages}{427}, \doiprefix\url{https://doi.org/10.3390/math7050427}
  (\bibinfo{year}{2019}).

\bibitem{Hernandez:2021}
\bibinfo{author}{Hernández, N.} \emph{et~al.}
\newblock \bibinfo{journal}{\bibinfo{title}{Retrieving the structure of
  probabilistic sequences of auditory stimuli from eeg data}}.
\newblock {\emph{\JournalTitle{Scientific Reports}}}
  \textbf{\bibinfo{volume}{11}}, \bibinfo{pages}{3520},
  \doiprefix\url{https://doi.org/10.1038/s41598-021-83119-x}
  (\bibinfo{year}{2021}).

\bibitem{Rissanen:1983}
\bibinfo{author}{Rissanen, J.}
\newblock \bibinfo{journal}{\bibinfo{title}{A universal data compression
  system}}.
\newblock {\emph{\JournalTitle{IEEE Transactions on Information Theory}}}
  \textbf{\bibinfo{volume}{29}}, \bibinfo{pages}{656--664},
  \doiprefix\url{https://doi.org/10.1109/TIT.1983.1056741}
  (\bibinfo{year}{1983}).

\bibitem{gkgame}
\bibinfo{author}{NeuroMat-FAPESP}.
\newblock \bibinfo{title}{The goalkeeper game}.
\newblock \bibinfo{howpublished}{https://game.numec.prp.usp.br/}
  (\bibinfo{year}{2022}).

\bibitem{Stern:2020}
\bibinfo{author}{Stern, R.~B.} \emph{et~al.}
\newblock \bibinfo{journal}{\bibinfo{title}{Goalkeeper game: A new assessment
  tool for prediction of gait performance under complex condition in people
  with parkinson's disease}}.
\newblock {\emph{\JournalTitle{Frontiers in Aging Neuroscience}}}
  \textbf{\bibinfo{volume}{0}},
  \doiprefix\url{https://doi.org/10.3389/fnagi.2020.00050}
  (\bibinfo{year}{2020}).

\bibitem{Hernandez:2023}
\bibinfo{author}{Hernández, N.}, \bibinfo{author}{Galves, A.},
  \bibinfo{author}{Garcia, J.}, \bibinfo{author}{Gubitoso, M.~D.} \&
  \bibinfo{author}{Vargas, C.~D.}
\newblock \bibinfo{journal}{\bibinfo{title}{Probabilistic prediction and
  context tree identification in the goalkeeper game}}.
\newblock {\emph{\JournalTitle{Arxiv}}}  (\bibinfo{year}{2023}).

\bibitem{smselection}
\bibinfo{author}{Galves, A.}, \bibinfo{author}{Leonardi, F.} \&
  \bibinfo{author}{Ost, G.}
\newblock \bibinfo{title}{Statistical model selection for stochastic systems
  with applications to bioinformatics, linguistics and neurobiology}.
\newblock
  \bibinfo{howpublished}{https://coloquio33.impa.br/pdf/33CBM15-eBook-preview.pdf}
  (\bibinfo{year}{2022}).

\bibitem{Hill:1982}
\bibinfo{author}{Hill, M.~A.} \& \bibinfo{author}{Dixon, W.~J.}
\newblock \bibinfo{journal}{\bibinfo{title}{Robustness in real life: A study of
  clinical laboratory data}}.
\newblock {\emph{\JournalTitle{International Biometric Society}}}
  \textbf{\bibinfo{volume}{38}}, \bibinfo{pages}{377--396}
  (\bibinfo{year}{1982}).

\bibitem{Rabbitt:1966}
\bibinfo{author}{Rabbitt, P.~M.}
\newblock \bibinfo{journal}{\bibinfo{title}{Errors and error correction in
  choice-response tasks}}.
\newblock {\emph{\JournalTitle{Journal of Experimental Psychology}}}
  \textbf{\bibinfo{volume}{71}}, \bibinfo{pages}{264--272},
  \doiprefix\url{https://doi.org/10.1037/h0022853} (\bibinfo{year}{1966}).

\bibitem{Rabbitt:1977}
\bibinfo{author}{Rabbitt, P.} \& \bibinfo{author}{Rodgers, B.}
\newblock \bibinfo{journal}{\bibinfo{title}{What does a man do after he makes
  an error? an analysis of response programming}}.
\newblock {\emph{\JournalTitle{Quarterly Journal of Experimental Psychology}}}
  \textbf{\bibinfo{volume}{29}}, \bibinfo{pages}{727--743},
  \doiprefix\url{https://doi.org/10.1080/1464074770840} (\bibinfo{year}{1977}).

\bibitem{Gehring:2001}
\bibinfo{author}{Gehring, W.~J.} \& \bibinfo{author}{Fencsik, D.~E.}
\newblock \bibinfo{journal}{\bibinfo{title}{Functions of the medial frontal
  cortex in the processing of conflict and errors}}.
\newblock {\emph{\JournalTitle{The Journal of Neuroscience: The Official
  Journal of the Society for Neuroscience}}} \textbf{\bibinfo{volume}{21}},
  \bibinfo{pages}{23},
  \doiprefix\url{https://doi.org/10.1523/JNEUROSCI.21-23-09430.2001}
  (\bibinfo{year}{2001}).

\bibitem{Jentzsch:2009}
\bibinfo{author}{Jentzsch, I.} \& \bibinfo{author}{Dudschig, C.}
\newblock \bibinfo{journal}{\bibinfo{title}{Why do we slow down after an error?
  mechanisms underlying the effects of posterror slowing}}.
\newblock {\emph{\JournalTitle{Quarterly journal of experimental psychology}}}
  \textbf{\bibinfo{volume}{62}}, \bibinfo{pages}{209--18},
  \doiprefix\url{https://doi.org/10.1080/17470210802240655}
  (\bibinfo{year}{2009}).

\bibitem{Eichele:2010}
\bibinfo{author}{Eichele, H.}, \bibinfo{author}{Juvodden, H.},
  \bibinfo{author}{Ullsperger, M.} \& \bibinfo{author}{Eichele, T.}
\newblock \bibinfo{journal}{\bibinfo{title}{Mal-adaptation of event-related eeg
  responses preceding performance errors}}.
\newblock {\emph{\JournalTitle{Frontiers in Human Neuroscience}}}
  \textbf{\bibinfo{volume}{4}}, \bibinfo{pages}{65},
  \doiprefix\url{https://doi.org/10.3389/fnhum.2010.00065}
  (\bibinfo{year}{2010}).

\bibitem{Danielmeier:2011}
\bibinfo{author}{Danielmeier, C.} \& \bibinfo{author}{Ullsperger, M.}
\newblock \bibinfo{journal}{\bibinfo{title}{Post-error adjustments}}.
\newblock {\emph{\JournalTitle{Frontiers in Psychology}}}
  \textbf{\bibinfo{volume}{2}}, \bibinfo{pages}{233},
  \doiprefix\url{https://doi.org/10.3389/fpsyg.2011.00233}
  (\bibinfo{year}{2011}).

\bibitem{Danielmeier:2011b}
\bibinfo{author}{Danielmeier, C.}, \bibinfo{author}{Eichele, T.},
  \bibinfo{author}{Forstmann, B.~U.}, \bibinfo{author}{Tittgemeyer, M.} \&
  \bibinfo{author}{Ullsperger, M.}
\newblock \bibinfo{journal}{\bibinfo{title}{Posterior medial frontal cortex
  activity predicts post-error adaptations in task-related visual and motor
  areas}}.
\newblock {\emph{\JournalTitle{The Journal of Neuroscience}}}
  \textbf{\bibinfo{volume}{31}}, \bibinfo{pages}{1780--1789},
  \doiprefix\url{https://doi.org/10.1523/JNEUROSCI.4299-10.2011.}
  (\bibinfo{year}{2011}).

\bibitem{Braem:2015}
\bibinfo{author}{Braem, S.}, \bibinfo{author}{Coenen, E.},
  \bibinfo{author}{Bombeke, K.}, \bibinfo{author}{Bochove, M. E.~V.} \&
  \bibinfo{author}{Notebaert, W.}
\newblock \bibinfo{journal}{\bibinfo{title}{Open your eyes for prediction
  errors}}.
\newblock {\emph{\JournalTitle{Cognitive Affective and Behavioral Neuroscience}}}
  \textbf{\bibinfo{volume}{15}}, \bibinfo{pages}{374--80},
  \doiprefix\url{https://doi.org/10.3758/s13415-014-0333-4}
  (\bibinfo{year}{2015}).

\bibitem{Wessel:2018}
\bibinfo{author}{Wessel, J.~R.}
\newblock \bibinfo{journal}{\bibinfo{title}{An adaptive orienting theory of
  error processing}}.
\newblock {\emph{\JournalTitle{Psychophysiology}}}
  \textbf{\bibinfo{volume}{55}}, \doiprefix\url{doi: 10.1111/psyp.13041}
  (\bibinfo{year}{2018}).

\end{thebibliography}

\section*{Acknowledgements}

This work is part of the activities of FAPESP  Research, Innovation and Dissemination Center for Neuromathematics (grant \# 2013/ 07699-0 , S.Paulo Research Foundation (FAPESP). 
This work is supported by CAPES (88882.33210 8/2019-01) and FAPESP (2022/00699-3) grants. 
A.G and C.D.V. were partially supported by CNPq fellowships (grants 314836/2021-7 and 310397/2021-9)
This article is  also supported by FAPERJ (\# E26/010002474/2016, \# CNE 202.785/2018 and \# E-26/010.002418/2019),  and FINEP ( \# 18.569-8) grants. 
The authors acknowledge the hospitality of the  Institut Henri Poincaré (LabEx CARMIN ANR-10-LABX-59-01) where part of this work was written.

\section*{Author contributions statement}

All authors have contributed equally to the article.

\section*{Additional information}

The data and the code of the algorithms used in the analysis are  available at the following repository: \url{https://github.com/PauloCabral-hub/Publications/tree/main/Passos_etal2023}. Instructions about the use of the algorithms are presented in README files included in the repository.
The authors have no competing interests as defined by Nature Research, or other interests that might be perceived to influence the interpretation of the article.


\pagebreak

\section*{Supplementary Material}

\begin{table}[h]
    \centering
 \begin{tabular}{ |c|c|c|c|c|c|c|c|c|c|c| } \hline  & \multicolumn{2}{|c|}{$w = 0$} & \multicolumn{2}{|c|}{$w = 01$} & \multicolumn{2}{|c|}{$w = 11$} & \multicolumn{2}{|c|}{$w = 21$} & \multicolumn{2}{|c|}{$w = 2$} \\ \hline $ID$ & $S$ & $F$ & $S$ & $F$ & $S$ & $F$ & $S$ & $F$ & $S$ & $F$ \\ \hline 1 & 0.254 & 0.265 & 0.593 & 0.555 & 0.256 & 0.281 & 0.198 & 0.212 & 0.295 & 0.466 \\  \hline 2 & 0.335 & 0.347 & 0.354 & 0.197 & 0.445 & 0.268 & 0.307 & 0.286 & 0.236 & 0.381 \\ 3 & 0.308 & 0.413 & 0.327 & 0.356 & 0.354 & 0.301 & 0.298 & 0.346 & 0.266 & 0.447 \\ 4 & 0.206 & 0.267 & 0.397 & 0.397 & 0.177 & 0.135 & 0.137 & 0.224 & 0.121 & 0.196 \\ 5 & 0.433 & 0.365 & 0.461 & 0.486 & 0.396 & 0.300 & 0.381 & 0.351 & 0.256 & 0.502 \\ 6 & 0.260 & 0.258 & 0.314 & 0.241 & 0.235 & 0.361 & 0.257 & 0.321 & 0.232 & 0.271 \\ 7 & 0.312 & 0.321 & 0.329 & 0.224 & 0.255 & 0.287 & 0.223 & 0.263 & 0.231 & 0.442 \\ 8 & 0.310 & 0.349 & 0.392 & 0.246 & 0.247 & 0.191 & 0.272 & 0.246 & 0.229 & 0.302 \\ 9 & 0.158 & 0.171 & 0.153 & 0.157 & 0.141 & 0.112 & 0.130 & 0.095 & 0.135 & 0.162 \\ 10 & 0.365 & 0.462 & 0.471 & 0.344 & 0.330 & 0.292 & 0.223 & 0.284 & 0.191 & 0.486 \\ 11 & 0.405 & 0.362 & 0.485 & 0.407 & 0.435 & 0.379 & 0.329 & 0.374 & 0.249 & 0.526 \\ 12 & 0.215 & 0.276 & 0.238 & 0.140 & 0.247 & 0.212 & 0.234 & 0.226 & 0.198 & 0.415 \\ 13 & 0.307 & 0.319 & 0.355 & 0.381 & 0.446 & 0.249 & 0.190 & 0.306 & 0.264 & 0.353 \\ 14 & 0.420 & 0.342 & 0.347 & 0.305 & 0.249 & 0.331 & 0.378 & 0.436 & 0.293 & 0.354 \\ 15 & 0.174 & 0.236 & 0.520 & 0.289 & 0.151 & 0.180 & 0.147 & 0.149 & 0.158 & 0.325 \\ 16 & 0.641 & 0.590 & 0.353 & 0.441 & 0.380 & 0.488 & 0.353 & 0.452 & 0.388 & 0.446 \\ 17 & 0.151 & 0.160 & 0.108 & 0.129 & 0.120 & 0.159 & 0.131 & 0.132 & 0.128 & 0.166 \\ 18 & 0.565 & 0.477 & 0.455 & 0.392 & 0.401 & 0.543 & 0.465 & 0.372 & 0.452 & 0.495 \\ 19 & 0.342 & 0.373 & 0.308 & 0.320 & 0.393 & 0.276 & 0.297 & 0.436 & 0.294 & 0.629 \\ 20 & 0.546 & 0.532 & 0.401 & 0.239 & 0.380 & 0.296 & 0.297 & 0.373 & 0.279 & 0.293 \\ 21 & 0.461 & 0.432 & 0.482 & 0.378 & 0.427 & 0.386 & 0.444 & 0.540 & 0.426 & 0.471 \\ 22 & 0.262 & 0.233 & 0.285 & 0.313 & 0.304 & 0.197 & 0.261 & 0.271 & 0.206 & 0.339 \\  \hline \end{tabular}
    \caption{Mean response times per participant and per context according to the prediction result at the last time the context 0 took place. ID indicates the participant number, $w$ the corresponding context, and $S$ and $F$, success and failure, respectively.}
    \label{tab:TS1}
\end{table}

\end{document}